\documentclass[preprint,amsmath,amssymb,aps,endfloats*]{revtex4-1}
\usepackage{graphicx}
\usepackage{color}
\usepackage{amsmath}
\usepackage{numprint}

\begin{document}
\title{Efficient numerical solver for first-principles transport calculation based on real-space finite-difference method}
\author{Shigeru Iwase}
\affiliation{Graduate School of Engineering, Osaka University, Suita, Osaka 565-0871, Japan}
\author{Takeo Hoshi}
\affiliation{Department of Applied Mathematics and Physics, Tottori University, Tottori, Tottori 680-8550, Japan}
\affiliation{JST-CREST, Kawaguchi, Saitama 332-0012, Japan}
\author{Tomoya Ono}
\affiliation{Center for Computational Sciences, University of Tsukuba, Tsukuba, Ibaraki 305-8577, Japan}
\affiliation{JST-PRESTO, Kawaguchi, Saitama 332-0012, Japan}
\begin{abstract}
We propose an efficient procedure to obtain Green's functions by combining the shifted conjugate orthogonal conjugate gradient (shifted COCG) method with the nonequilibrium Green's function (NEGF) method based on a real-space finite-difference (RSFD) approach. The bottleneck of the computation in the NEGF scheme is matrix inversion of the Hamiltonian including the self-energy terms of electrodes to obtain perturbed Green's function in the transition region. This procedure first computes unperturbed Green's functions and calculates perturbed Green's functions from the unperturbed ones using a mathematically strict relation. Since the matrices to be inverted to obtain the unperturbed Green's functions are sparse, complex-symmetric and shifted for a given set of sampling energy points, we can use the shifted COCG method, in which once the Green's function for a reference energy point has been calculated, the Green's functions for the other energy points can be obtained with a moderate computational cost. We calculate the transport properties of a C$_{60}$@(10,10) carbon nanotube (CNT) peapod suspended by (10,10)CNTs as an example of a large-scale transport calculation. The proposed scheme opens the possibility of performing large-scale RSFD-NEGF transport calculations using massively parallel computers without the loss of accuracy originating from the incompleteness of the localized basis set.
\end{abstract}

\maketitle

\section{Introduction}
\label{sec:1}
Since the discovery of quantized conductance in atomic-size contacts, there has been considerable interest in studying the electronic transport in nanoscale systems. So far, a number of calculation methods to investigate the electron transport properties through a nanostructure embedded by semi-infinite electrodes have been proposed such as the nonequilibrium Green's function (NEGF) method \cite{negf}, recursion transfer matrix method \cite{rtm}, Lippmann-Schwinger method \cite{ls} and overbridging boundary-matching (OBM) method \cite{book,obm}. Among them, the NEGF approach is most commonly used in connection with tight-binding models and first-principles methods for solving transport problems. In the NEGF formalism, localized basis sets consisting of either atomic orbitals or Gaussians are extensively utilized because it is straightforward to partition a whole system into a transition region and electrodes. However, the incompleteness of the basis sets sometimes degrades the accuracy \cite{ghost} and the diffuse functions in the basis sets prevent us from achieving high-performance computing on massively parallel computers.

On the other hand, the calculation method using the real-space finite-difference (RSFD) approach~\cite{book,cheliko,tsdg} does not suffer from these problems. However, the linear-equation solver used to obtain the perturbed Green's functions of a transition region becomes computationally intensive as the numbers of grid points and sampling energy points increase. More efficient linear-equation solvers are required to perform large-scale first-principles transport calculations by the NEGF approach using the RSFD scheme.

In this study, we propose an efficient scheme to compute perturbed Green's functions that combines the procedure developed in the OBM method and the shifted conjugate orthogonal conjugate gradient (COCG) algorithm \cite{sCOCG,ss}. In the NEGF method, the matrix to be inverted includes the self-energy terms of electrodes, which are energy-dependent and destroy the shift invariance of the Krylov subspace with respect to the energy of electrons. The total computational cost of obtaining perturbed Green's functions is proportional to the number of energy points. On the other hand, unperturbed Green's functions are first calculated by the OBM method and then the effect of the electrodes is introduced in the wavefunction-matching procedure.
Since the matrix to be inverted does not include nonlinear energy dependent terms, the unperturbed Green's functions are obtained by solving a set of shifted linear equations with respect to the energy. The shifted linear equations can be solved by the shifted COCG method, in which the calculation is actually carried out only for a reference energy point and the solutions for the other shifted energy points are obtained with a moderate computational cost owing to the shift invariance of the Krylov subspace.
Employing the mathematically strict relationship between perturbed and unperturbed Green's functions developed in the OBM method \cite{rsfd-negf}, one can greatly decrease the computational cost of the linear-equation solver.

The rest of this paper is organized as follows. In Sec.~\ref{sec:2}, the NEGF method using the RSFD approach is briefly introduced. In Sec.~\ref{sec:3}, we state the problem of obtaining perturbed Green's functions and introduce the computational scheme combining the procedure developed in the OBM method and the shifted COCG algorithm together with a numerical example. In Sec.~\ref{sec:4}, to demonstrate the efficiency and applicability of the newly developed method, we present a transport calculation of a C$_{60}$@(10,10) carbon nanotube (CNT) peapod connected to CNT electrodes. Finally, we summarize our work in Sec.~\ref{sec:5}.

\section{Computational scheme}
\label{sec:2}
\subsection{Hamiltonian of whole system}
\begin{figure}[h]
\includegraphics{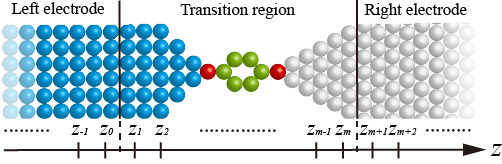}
\caption{Schematic of transport calculation model. The transition region is suspended between the left and right semi-infinite electrodes.}
\label{fig:junction}
\end{figure}
We consider ballistic transport through a nanostructure sandwiched between two semi-infinite crystalline electrodes as shown in Fig. \ref{fig:junction}. The periodic boundary condition is applied in the $x$ and $y$ directions, while the semi-infinite boundary condition is imposed in the $z$ direction. The system under investigation can be divided into three regions: the left electrode (L), the transition region (T) and the right electrode (R).  By introducing localized basis sets or real-space grids, we can decompose a whole Hamiltonian $\hat{H}$ into the following matrix form:
\begin{equation}
\label{eqn: H matrix}
\hat{H}=
\left
[\begin{array}{c|c|c}
\hat{H}_L&\hat{B}_{LT}&0 \\ \hline
\hat{B}_{LT}^{\dagger}&\hat{H}_{T} &\hat{B}_{TR} \\ \hline
     0  & \hat{B}_{TR}^{\dagger}&\hat{H}_{R}
\end{array}
\right],
\end{equation}
where $\hat{H}_L$, $\hat{H}_T$ and $\hat{H}_R$ are the Hamiltonian matrices for the left electrode, transition region and right electrode, respectively. $\hat{B}_{LT}$ ($\hat{B}_{TR}$) is the coupling term between the transition region and the left (right) electrode. Because of electronic screening effects, the coupling between the left and right electrodes is zero when the transition region is sufficiently large.

We next derive the Kohn-Sham (KS) Hamiltonian \cite{ks} for the RSFD scheme. Since the direction of current flow is assumed to be along the $z$ axis, the KS Hamiltonian is composed of the block-matrix elements with a dimension of $N_{xy}(=N_x\times{N_y})$, where $N_{x}$ and $N_{y}$ are the numbers of grid points in the $x$ and $y$ directions, respectively. Wavefunction values on the $x$-$y$ plane at the point $z = z_k$ are expressed by the column vector ${\Psi(z_k)}$ with $N_{xy}$ components given by \{$\psi(x_i,y_j,z_k):i=1,\ldots,N_x,j=1,\ldots,N_y$\}. To simplify the explanation, we here restrict ourselves to the case of the central finite-difference approximation, i.e., $N=1$ in Eq.~(1) of Ref.~\onlinecite{cheliko}, and local pseudopotentials. The extensions to higher-order finite-difference approximations and the inclusion of nonlocal parts of pseudopotentials are given in Ref.~\onlinecite{book}. The KS equation for a whole system is expressed as a relation between three adjacent terms along the $z$ direction
\begin{equation}
-B_z^{\dagger}\Psi(z_{k-1})+[E-A(z_k)]\Psi(z_k)-B_z\Psi(z_{k+1})=0 \;\;(k = -\infty, \ldots, -1, 0, 1, \ldots, \infty),
\label{eqn:sabun1}
\end{equation}
where $E$ is the energy of an electron, $A(z_k)$ is an $N_{xy}$-dimensional block tridiagonal matrix including the potentials on the $x$-$y$ plane at $z=z_{k}$ and the coefficients of the finite-difference approximation, and $B_z$ is a constant matrix that is proportional to the $N_{xy}$-dimensional unit matrix $I$.
The discretized KS equation of the whole system is written as
\begin{equation}
\label{eqn:matrix eq.}
[E-\hat{H}]
\left[
\begin{array}{c}
\vdots  \\ \Psi(z_0) \\ \Psi(z_1)  \\ \vdots  \\ \Psi(z_m) \\ \Psi(z_{m+1}) \\ \vdots
\end{array}
\right]
=0,
\end{equation}
where $\hat{H}$ is the block tridiagonal matrix
\begin{equation}
\label{eqn:matrix eq.}
\hat{H}=
\left
[\begin{array}{ccc|ccccc|ccc}
\ddots & \ddots& 0&&&&&&&& \\
\ddots & A_{z_{-2}} &B_{z}        &&&&&&&  \\ 
       0  &     B_{z}^{\dagger}    & A_{z_{-1}}   & B_{z_{1}}     &&&&&&&  \\ \hline
&&                         B_{z}^{\dagger}        & A_{z_0} &B_{z}  &&&&&&  \\
&&&                                      B_{z}^{\dagger}     &A_{z_1} &B_{z}  &&&&& \\
&&&&                                      \ddots & \ddots & \ddots   &&&&      \\
&&&&&                                       B_{z}^{\dagger} & A_{z_{m}}&B_{z}   &&& \\
&&&&&&                                               B_{z}^{\dagger}          & A_{z_{m+1}} & B_{z}   && \\ \hline
&&&&&&&                                                               B_{z}^{\dagger}    & A_{z_{m+2}} & B_{z} &              0 \\
&&&&&&&&                                                                         B_{z}^{\dagger}           & A_{z_{m+3}} & \ddots \\
&&&&&&&&                                                                          0            &      \ddots      & \ddots  \\            
\end{array}
\right]
\end{equation}
with $A_{z_{k}}=A(z_{k})$. The partitioning in Eq. (\ref{eqn:matrix eq.}) corresponds to those in Fig. \ref{fig:junction} and Eq. (\ref{eqn: H matrix}). Since the KS Hamiltonian matrix for the RSFD scheme is very sparse and block tridiagonal, we can employ efficient iterative algorithms such as the steepest descent (SD) or conjugate gradient (CG) method to solve the eigenvalue and matrix inversion problems.
\subsection{NEGF method using RSFD approach}
This subsection gives a brief explanation of the NEGF method for later convenience. The NEGF method gives the electron transport properties without calculating the scattering wavefunction explicitly. The Green's function of a whole system can be defined as the resolvent of the Hamiltonian matrix of Eq. (\ref{eqn:matrix eq.}),
\begin{equation}
\label{eqn:GF1}
\hat{G}(Z)=
\left[
\begin{array}{c}
Z-\hat{H}
\end{array}
\right]^{-1},
\end{equation}
where $Z$($=E+i\eta$) is a complex energy with $\eta$ being a positive infinitesimal. The important quantity determining electron transport properties is the perturbed Green's function of the transition region written as
\begin{equation}
\label{eqn:GF2}
\hat{G}_T(Z)=
\left[
\begin{array}{c}
Z-\hat{H}_T-\hat{\Sigma}_L(Z)-\hat{\Sigma}_R(Z)
\end{array}
\right]^{-1},
\end{equation}
where 
\begin{equation}
\label{eqn:self-energy}
\hat{\Sigma}_L(Z)=\hat{B}_{LT}^{\dagger}\hat{\mathcal{G}}_L(Z)\hat{B}_{LT} \:\: \mbox{and}\:\: \hat{\Sigma}_R(Z)=\hat{B}_{TR}\hat{\mathcal{G}}_R(Z)\hat{B}_{TR}^{\dagger}
\end{equation}
are the self-energy terms of the left and right electrodes, respectively. Here
\begin{equation}
\label{eqn:surface green's function}
\hat{\mathcal{G}}_L(Z)=[Z-\hat{H}_{L}]^{-1}  \:\: \mbox{and}\:\: \hat{\mathcal{G}}_R(Z)=[Z-\hat{H}_{R}]^{-1}
\end{equation}
are the surface Green's functions of the semi-infinite left and right electrodes, respectively. In the RSFD-NEGF scheme, since $\hat{B}_{LT}$ ($\hat{B}_{TR}$) has only one nonzero $N_{xy}$-dimensional block-matrix element $B_z$ [see Eq.~(\ref{eqn:matrix eq.})], the self-energy terms are found to take the very simple forms of 
\begin{eqnarray}
{\textstyle \hat{\Sigma}_L}(Z) & = &
\left[ 
\begin{array}{cccc}
{\textstyle \Sigma_L}(z_0;Z) & 0 & \cdots & 0 \\ 
0 & 0 & \cdots & 0 \\ 
\vdots &  & \ddots &  \\ 
0 & 0 & \cdots & 0 
\end{array}
\right]
\end{eqnarray}
and
\begin{eqnarray}
{\textstyle \hat{\Sigma}_R}(Z) & = &
\left[ 
\begin{array}{cccc}
0 & \cdots & 0 & 0 \\ 
  & \ddots &   & \vdots   \\ 
0 & \cdots & 0 & 0 \\ 
0 & \cdots & 0 & {\textstyle \sum _R}(z_{m+1};Z) 
\end{array}
\right],
\label{eqn:sec2-01}
\end{eqnarray}
where
\begin{eqnarray}
{\textstyle \Sigma_L}(z_0;Z) & = & B_z^{\dagger} \mathcal{G}_L(z_{-1},z_{-1};Z) B_z 
\end{eqnarray}
and
\begin{eqnarray}
{\textstyle \Sigma_R}(z_{m+1};Z) & = & B_z \mathcal{G}_R(z_{m+2},z_{m+2};Z) B_z^{\dagger}
\label{eqn:sec2-02}
\end{eqnarray}
with $\mathcal{G}_{\{L,R\}}(z_k,z_l;Z)$ being the $N_{xy}$-dimensional $(k,l)$ block-matrix element of $\hat{\mathcal{G}}_{\{L,R\}}(Z)$. The perturbed Green's function of the transition region $\hat{G}_T(Z)$ can be obtained by inverting the finite-size matrix given as
\begin{equation}
\label{eqn:NEGF Hamiltonian}
\left[
\begin{array}{ccccc}
D({z_0};Z)-{\textstyle \Sigma_L}(z_0;Z)       & -B_z                    &&&0 \\ 
-B_z^{\dagger} &D({z_1};Z)&-B_z      &&  \\
&\ddots               &\ddots   &\ddots& \\
&&                            -B_{z}^{\dagger}&D({z_m};Z)&-B_{z} \\
0&&                                                    &-B_{z}^{\dagger}&D({z_{m+1}};Z)-{\textstyle \Sigma_R}(z_{m+1};Z)\\ 
\end{array}
\right],
\end{equation}
where $D({z_l};Z)=Z-A(z_l)$. 

The expression for the conductance at a real energy $E$ is given by the Fisher-Lee formula \cite{fisherlee}:

\begin{equation}
\label{eqn:TF}
T(E)=
\mathrm{Tr}\left[
\begin{array}{c}
\Gamma_L(z_0;E)G_T^a(z_0,z_{m+1};E)\Gamma_R(_{m+1};E)G_T^r(z_{m+1},z_0;E)
\end{array}
\right],
\end{equation}
where $G_T^r(E)$ is the retarded Green's function of the transition region defined as
\begin{equation}
G_T^r(z_k,z_l;E)=\lim_{\eta \to 0^+}G_T(z_k,z_l;E+i\eta),
\end{equation}
$G_T^a(E)(=[G_T^r(E)]^{\dagger})$ is the advanced Green's function, and $\Gamma_{L(R)}$ is the coupling matrix calculated from the imaginary part of the self-energy term,
\begin{equation}
\Gamma_{X}(z_k;E)=i[\Sigma_X(z_k;E)-\Sigma_X^{\dagger}(z_k;E)] \;\;\;\;\ (X\in{L,R}).
\end{equation}


\section{Efficient procedure to obtain perturbed Green's function}
\label{sec:3}
\subsection{Difficulties in obtaining perturbed Green's function by iterative method}
The simplest way to obtain $\hat{G}_T(Z)$ is the direct inversion of the matrix given by Eq. (\ref{eqn:NEGF Hamiltonian}). However, direct matrix inversion becomes impractical when the matrix size increases. In addition, iterative solvers for linear equations, such as the SD and CG methods, are not efficient in this case because the matrix to be inverted given by Eq. (\ref{eqn:NEGF Hamiltonian}) is neither Hermitian nor very sparse owing to the self-energy terms. In addition, the a computation of $\hat{G}_T(Z)$, which is carried out independently at each energy, is a computationally demanding task. To circumvent these difficulties, we propose a computational procedure employing the unperturbed Green's function used in the OBM method \cite{obm,book} and the shifted COCG algorithm \cite{sCOCG,ss} that uses the collinear residual theorems \cite{frommer} in the following subsection.
\subsection{Unperturbed Green's function and shifted COCG method}
The unperturbed Green's function of the transition is written as
\begin{equation}
\label{eqn:GF4}
\hat{\mathcal{G}}_T(Z)=[Z-\hat{H}_{T}]^{-1}.
\end{equation}
The relationship between the unperturbed and perturbed Green's functions is expressed as the following equation, which was derived in Ref. \onlinecite{rsfd-negf}.
\begin{equation}
\left[ 
\begin{array}{c}
G_T(z_0,z_l;Z) \\ 
G_T(z_1,z_l;Z) \\ 
\\
\vdots                 \\
G_T(z_l,z_l;Z) \\
\vdots                 \\
\\
G_T(z_m,z_l;Z) \\ 
G_T(z_{m+1},z_l;Z) \\ 
\end{array}
\right]
=\hat{\mathcal{G}}_T(Z) \left[ 
\begin{array}{c}
\sum_L(z_0;Z)G_T(z_0,z_l;Z) \\ 
0 \\ 
\vdots                 \\ 
0 \\
I  \\
0 \\ 
\vdots                 \\ 
0 \\
\sum_R(z_{m+1};Z)G_T(z_{m+1},z_l;Z) \\ 
\end{array}
\right]
\begin{array}{l}
 \\ 
 \\ 
 \\ 
 \\ 
.\leftarrow \mbox{the} \hspace{2mm} l\mbox{th} \\
 \\ 
 \\ 
 \\ 
 \\ 
\end{array}
\label{eqn:sec3-01a}
\end{equation}
For $l=0$ and $m+1$,
\begin{eqnarray}
G_T(z_0,z_0;Z) &=& \tilde{\mathcal{G}}_T(z_0,z_0;Z)\left[ I-{\textstyle \sum_L}(z_0;Z)\tilde{\mathcal{G}}_T(z_0,z_0;Z) \right]^{-1},\label{eqn:sec3-02}
\\
G_T(z_{m+1},z_0;Z) &=& \Bigl[I - \mathcal{G}_T(z_{m+1},z_{m+1};Z){\textstyle \sum_R}(z_{m+1};Z)\Bigr]^{-1}  \nonumber \\
&&  \times\ \mathcal{G}_T(z_{m+1},z_{0};Z) \left[ I-{\textstyle \sum_L}(z_0;Z)\tilde{\mathcal{G}}_T(z_0,z_0;Z) \right]^{-1},  \\
G_T(z_0,z_{m+1};Z) &=& \Bigl[ I - \mathcal{G}_T(z_{0},z_{0};Z){\textstyle \sum_L}(z_{0};Z)\Bigr]^{-1}\mathcal{G}_T(z_{0},z_{m+1};Z) \nonumber \\ 
& & \times \left[ I-{\textstyle \sum_R}(z_{m+1};Z)\tilde{\mathcal{G}}_T(z_{m+1},z_{m+1};Z) \right]^{-1}, 
\end{eqnarray}
and
\begin{eqnarray}
G_T(z_{m+1},z_{m+1};Z) &=& \tilde{\mathcal{G}}_T(z_{m+1},z_{m+1};Z)\left[ I-{\textstyle \sum_R}(z_{m+1};Z)\tilde{\mathcal{G}}_T(z_{m+1},z_{m+1};Z) \right]^{-1}, 
\label{eqn:sec3-03}
\end{eqnarray}
where $\tilde{\mathcal{G}}_T(z_i,z_j;Z)$ is the modified $\mathcal{G}_T(z_i,z_j;Z)$ under the influence of the self-energy terms, ${\textstyle \sum_R}$ and ${\textstyle \sum_L}$, which is expressed as
\begin{eqnarray}
\tilde{\mathcal{G}}_T(z_0,z_0;Z) &=& \mathcal{G}_T(z_0,z_0;Z) \nonumber \\
&& \hspace{-20mm} + \mathcal{G}_T(z_0,z_{m+1};Z) {\textstyle \sum_R}(z_{m+1};Z) \nonumber \\
&& \hspace{-20mm} \times\ \Bigl[ I - \mathcal{G}_T(z_{m+1},z_{m+1};Z){\textstyle \sum_R}(z_{m+1};Z)\Bigr]^{-1}\mathcal{G}_T(z_{m+1},z_{0};Z)
\end{eqnarray}
and
\begin{eqnarray}
\tilde{\mathcal{G}}_T(z_{m+1},z_{m+1};Z) &=& \mathcal{G}_T(z_{m+1},z_{m+1};Z) \nonumber \\
&& \hspace{-20mm} + \mathcal{G}_T(z_{m+1},z_{0};Z) {\textstyle \sum_L}(z_0;Z) \nonumber \\
&& \hspace{-20mm} \times\ \Bigl[ I - \mathcal{G}_T(z_{0},z_{0};Z){\textstyle \sum_L}(z_{0};Z)\Bigr]^{-1} \mathcal{G}_T(z_{0},z_{m+1};Z). 
\label{eqn:sec3-06}
\end{eqnarray}
The advantage of computing the unperturbed Green's function is that the matrix inverted in Eq. (\ref{eqn:GF4}), $Z-\hat{H}_T$, is very sparse and complex symmetric. Moreover, $Z_k-\hat{H}_T$ are shifted matrices for a given set of energy points $Z_k$ $(k=1,2,...)$, which is in contrast to $Z_k-\hat{H}_T-\hat{\Sigma}_L(Z_k)-\hat{\Sigma}_R(Z_k)$ in Eq.~(\ref{eqn:GF2}). Therefore, shifted methods such as the shifted COCG method \cite{sCOCG} can be applied to obtain unperturbed Green's functions.

To solve the complex-symmetric linear equations $(Z_k-\hat{H}_T)x^{(k)}=e_j $, where $e_{j}$ denotes the $j$th unit vector, by the conventional COCG method \cite{COCG}, we build up the Krylov subspaces independently at each complex energy $Z_k$, which requires time-consuming matrix-vector operations for each energy. On the other hand, the shifted COCG method makes use of the fact that Krylov subspaces are shift-invariant; to solve the linear equations for a given set of sampling energy points, the spanned Krylov subspace is common to all the energy points. Once the Krylov subspace has been built up for the reference energy point, we can obtain the Green's functions for all the sampling energy points without matrix-vector operations during the iteration steps.

It is noteworthy that we do not need to store the whole matrices of the unperturbed Green's functions $\hat{\mathcal{G}}_T(Z)$ in memory. To calculate the transport properties by Eq. (\ref{eqn:TF}), the block-matrix elements of the perturbed Green's functions, $G_T(z_{m+1},z_0;Z)$ and $G_T(z_0,z_{m+1};Z)$, are required. In addition, the density of states is relevant to the diagonal elements of the perturbed Green's functions, $G_T(z_{l},z_{l};Z)$. The evaluation of $G_T(z_{l},z_{l};Z)$ requires the additional computation of the off-diagonal block elements $G_T(z_0,z_{l};Z)$ and $G_T(z_{m+1},z_l;Z)$, which are analytically determined in an analogous manner to the derivation of Eqs. (\ref{eqn:sec3-02})-(\ref{eqn:sec3-03}) from Eq. (\ref{eqn:sec3-01a}). It is sufficient to store a couple of block-matrix elements of the unperturbed Green's functions so as to perform matrix-matrix operations between the $N_{xy}$-dimensional block-matrix elements of the Green's functions and the self-energy terms. Thus, we can obtain the perturbed Green's functions with a moderate computation time and memory load. 

There is no guarantee that the residual norms of all the sampling energy points satisfy the convergence criterion when that of the reference energy point is sufficiently small. Because the iterations are performed until all the residual norms satisfy the convergence criterion, the total number of iterations increases, resulting in the slight increase of CPU time in the shifted COCG method with respect to the number of sampling energy points. In addition, when the norm of the residual vector of the seed system becomes small, the numerical precision of the residual vectors for the sampling energy points degrades. The seed switching technique \cite{ss}, which switches the reference energy point to the point where the residual norm is largest among the sampling energy points after the solution of the reference energy point converges, enables us to build up the residual vectors for the sampling energy points without loss of numerical precision and significant increase of the computational cost. The detail of the seed switching technique is introduced in Appendix.

\subsection{Numerical test}
To demonstrate the efficiency of the present scheme, we apply the shifted COCG method to the calculation of the unperturbed Green's functions of the transition region for a Na atomic wire. The valence electron-ion interaction is described using norm-conserving pseudopotentials \cite{norm} generated by the scheme proposed by Troullier and Martins \cite{tm}. Exchange-correlation effects are treated within the local density approximation \cite{lda} of density functional theory \cite{dft,ks}. The central finite-difference approximation is adopted for the second-order derivation arising from the kinetic energy operator in the KS equation. The dimensions of the supercell are $20.0\times20.0\times7.0$ bohr and a single Na atom is contained in the supercell. The real-space grid spacing is chosen to be $\sim0.45$ bohr, which corresponds to a total Hamiltonian dimension of 12800. All calculations are carried out on a workstation with an Intel Xeon E5-2667v2 CPU.
\begin{figure}[h]
\centering
\includegraphics{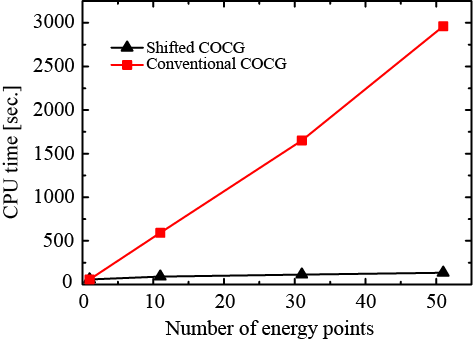}
\caption{CPU time required to obtain unperturbed Green's functions of Na atomic wire. The red squares are the results obtained by the conventional COCG method and the black triangles are those obtained by the shifted COCG method. The energy points are shifted from the Fermi energy by 0.1 eV.}
\label{fig:calctime}
\end{figure}
Figure \ref{fig:calctime} shows the CPU time versus the number of sampling energy points for the shifted COCG and conventional COCG methods. The initial reference energy point is chosen to be the Fermi energy. For the conventional COCG method, the CPU time is proportional to the number of energy points because the Green's functions are calculated independently at each energy point. On the other hand, in the shifted COCG method, the matrix-vector operations are carried out only at the reference energy point and the cost of the scalar-vector operations to update the approximate solutions for all the shifted energy points is negligibly small. The increase in the computational time is attributed to the time-consuming matrix-vector operations in the additional iteration steps to reach global convergence.

\section{Application}
\label{sec:4}
\begin{figure}[h]
\includegraphics{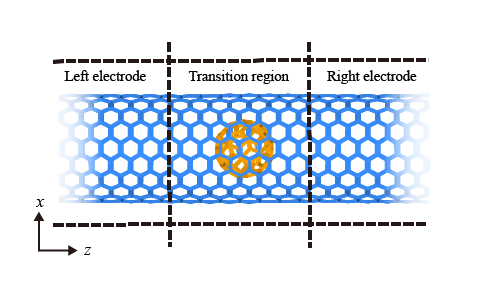}
\caption{Atomic structure of (10,10)CNT with single C$_{60}$ in transition region. The dashed lines are the boundaries of the supercell used to determine the atomic structure and KS effective potential.}
\label{fig:peapod}
\end{figure}
We next compute the conductance spectrum of a C$_{60}$@(10,10)CNT peapod, in which a (10,10)CNT encapsulates a C$_{60}$ molecule, as an application of the proposed method to a large system. Figure \ref{fig:peapod} shows the computational model. This system has been observed by transmission electron microscopy \cite{peapod3} and its electronic structure has been intensively studied by theoretical calculations \cite{peapod,peapod2,ohno}. However, the transport properties of peapods have mostly been investigated using the tight-binding approximation \cite{ohno} because of the large system size required for calculations.

The size of the transition region is taken to be $42.2\times42.2\times32.6$ bohr and the transition region contains 340 carbon atoms. The initial carbon-carbon distance is set to 2.68 bohr \cite{peapod2} and structural relaxation is carried out. To determine the KS effective potential for the transport calculation, the conventional supercell indicated by the rectangle in Fig. \ref{fig:peapod} is used, where the periodic boundary condition in imposed in all directions, and the grid spacing is $\sim0.45$ bohr, which gives a matrix of $774144\times774144$ for the Hamiltonian of the transition region. The number of sampling energy points in the transport calculation is 101. Calculations are implemented using 64 Intel Xeon E5-2697v2 CPUs. The total elapsed time required to calculate the unperturbed Green's functions for 101 energy points is 15294 second, in which the average number of iterations required to satisfy the convergence criterion is 6713, while the elapsed time and the average number of iterations are 6029 second and 3122, respectively, when the unperturbed Green's function is only calculated for the Fermi energy.

\begin{figure}[h]
\begin{center}
\includegraphics{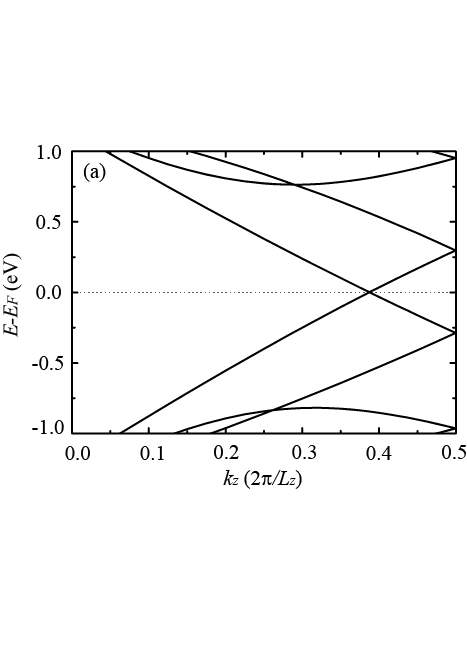}
\\
\includegraphics{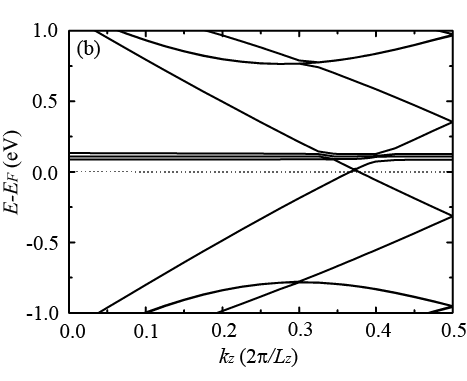}
\end{center}
\caption{Electronic band structures of (a) (10,10)CNT and (b) C$_{60}$@(10,10)CNT. The Fermi level is marked by the dotted line.}
\label{fig:bands}
\end{figure}

The electronic band structures of the (10,10)CNT and C$_{60}$@(10,10)CNT peapod are shown in Figs.~\ref{fig:bands}(a) and \ref{fig:bands}(b), respectively. In the band structure of the C$_{60}$@(10,10)CNT peapod, there are three flat bands above the Fermi level, which originate from the $t_{1u}$ orbitals of the isolated C$_{60}$ molecule. Owing to the hybridization between the $t_{1u}$ orbitals of the C$_{60}$ molecule and $\pi$ orbitals of the (10,10)CNT, the $\pi$ bands of the (10,10)CNT are divided by the bands originating from the $t_{1u}$ orbitals.

\begin{figure}[h]
\includegraphics{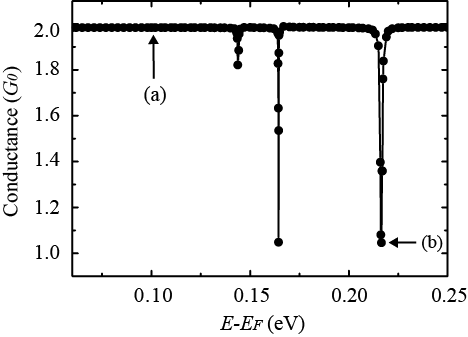}
\caption{Conductance spectrum of C$_{60}$@(10,10)CNT.}
\label{fig:spectra}
\end{figure}

\begin{figure}[h]
\begin{center}
\includegraphics{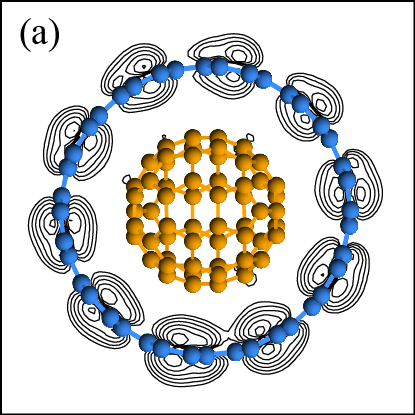}
\\
\includegraphics{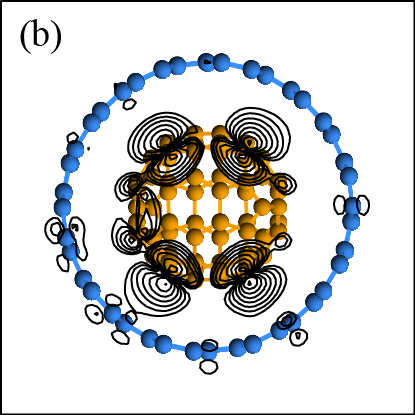}
\end{center}
\caption{Charge density distributions of scattering wavefunctions of C$_{60}$@(10,10)CNT. (a) and (b) correspond to the energies indicated by the arrows in Fig. \ref{fig:spectra}. The spheres represent the positions of carbon atoms. Each contour represents twice or half the charge density of the adjacent contour lines. The lowest-density contour represents a density of $5.0\times10^{-4}$ e/$\AA^3$}
\label{fig:swf}
\end{figure}

The conductance spectrum of the C$_{60}$@(10,10)CNT is plotted in Fig. \ref{fig:spectra}. Owing to the use of the shifted COCG method, we can sample a large number of energy points so as to identify the spiky dips in the spectrum. It is found that three dips appear near the energy range of the flat bands. Figure \ref{fig:swf} shows a real-space picture of the charge densities of scattering wavefunctions, which is useful for understanding the spatial behaviors of the transport phenomenon. The charge density distribution with the energy at $T(E)\approx2$ spreads around the (10,10)CNT, while that with the energy at the dip of the conductance spectrum accumulates in the vicinity of the C$_{60}$ molecule. These results imply the incident electrons are scattered by the $t_{1u}$ orbitals of the encapsulated C$_{60}$ molecule. Since peapods have recently become experimentally accessible structures, this information should be helpful in understanding and predicting the resonant scattering in experimental transport measurements \cite{sts}.

\section{Conclusion}
\label{sec:5}
An efficient scheme to compute perturbed Green's functions, which combines the procedure developed in the OBM method based on the RSFD approach and the shifted COCG algorithm, has been proposed. Since the number of grid points in the RSFD approach is larger than the number of bases in the tight-binding approach, the computational cost of obtaining the perturbed Green's functions of the transition region has been the bottleneck. In this procedure, unperturbed Green's functions are calculated and perturbed Green's functions are obtained by employing the mathematically strict relationship between the perturbed and unperturbed Green's functions. The main advantage of the present scheme for computing unperturbed Green's functions is that the spanned Krylov subspace is common to all the sampling energy points. Therefore, once the Krylov subspace for the reference energy point has been built up, the Green's functions for all the sampling energy points can be obtained without time-consuming matrix-vector operations. To demonstrate the potential power of the present method, the transport properties of a C$_{60}$@(10,10)CNT peapod suspended between (10,10)CNTs are investigated. By sampling a large number of energy points owing to the use of the present method, the scattering attributed to the coupling between the states of the CNT and C$_{60}$ can be observed in the calculated conductance spectrum.

The present technique enables us to efficiently obtain perturbed Green's functions in the framework of the RSFD approach. The RSFD scheme of first-principles calculations is free of problems concerning the completeness of the basis set that appear in the methods using localized basis sets of either atomic orbitals or Gaussians. Moreover, this scheme has the advantage of being scalable to massively parallel computers without compromising on precision. We believe that this development will be important for executing large-scale transport calculations using massively parallel computers.
\section*{Acknowledgment}
The authors would like to thank Professor Shao-Liang Zhang of Nagoya University, Professor Tomohiro Sogabe of Aichi Prefectural University, and Professor Susumu Yamamoto of Tokyo University of Technology for contributing to our fruitful discussions. This research was partially supported by the Computational Materials Science Initiative (CMSI) and a Grant-in-Aid for Scientific Research on Innovative Areas (Grant No. 22104007) from the Ministry of Education, Culture, Sports, Science and Technology, Japan. The numerical calculations were carried out using the computer facilities of the Institute for Solid State Physics at the University of Tokyo, the Center for Computational Sciences at University of Tsukuba, and K computer under a trial use at Advanced Institute for Computational Science at RIKEN.

\appendix*
\section{Shifted COCG algorithm and seed switching technique}
The COCG method is a numerical method for complex-symmetric linear equations as
\begin{equation}
Ax=b, \label{eqn:Eq}\\
\end{equation}
where A is not Hermitian but complex symmetric, $A=A^T \neq A^H$. Its algorithm gives the vectors $x_{n}$, $p_{n}$ and $r_{n}$ of the $n$th iteration as follows under the initial conditions, $x_{0}=p_{0}=0$, $r_{-1}=r_{0}=b$, $\alpha_{-1}=1$, and $\beta_{-1}=0$ \cite{ss}.
\begin{eqnarray}
x_{n}&=&x_{n-1}+\alpha_{n-1}p_{n-1},      \label{eqn:xvec}         \\ 
r_{n}&=&(1+\frac{\beta_{n-2}\alpha_{n-1}}{\alpha_{n-2}}-\alpha_{n-1}A)r_{n-1}-\frac{\beta_{n-2}\alpha_{n-1}}{\alpha_{n-2}}r_{n-2},      \label{eqn:rvec}         \\
p_{n}&=&r_{n}+\beta_{n-1}p_{n-1},                    \label{eqn:pvec}        \\
\alpha_{n-1}&=&\frac{(r_{n-1},r_{n-1})}{(r_{n-1},Ar_{n-1})-\frac{\beta_{n-2}}{\alpha_{n-2}}(r_{n-1},r_{n-1})}, \label{eqn:alpha} \\  
\beta_{n-1}&=&\frac{(r_{n},r_{n})}{(r_{n-1},r_{n-1})}. \label{eqn:beta}
\end{eqnarray}

Now we solve $m$ sets of the shifted linear equations $(A+\sigma_i I)x^{(i)}=b$ using the reference system $(A+\sigma_s I)x=b$, where $\sigma_i $ $(i=1,2,\ldots,m)$ is a given set of the sampling energy points and $I$ is an identity matrix. 
It is proved that the residual vector $r_n^{(i)}$ for the sampling energy point and the residual vector $r_n$ for the reference system are collinear:
\begin{equation}
\label{eqn:collinear}
r_n^{(i)}=\frac{1}{\pi_n^{(s,i)}}r_n, \\
\end{equation}
where $\pi_n^{(s,i)}$ is a scalar variable \cite{frommer}. Using the collinear relation of Eq.~(\ref{eqn:collinear}), the algorithm of the shifted COCG method is given as follows.
\begin{eqnarray}
\pi_{n+1}^{(s,i)}&=&(1+\frac{\beta_{n-1}\alpha_{n}}{\alpha_{n-1}}+\alpha_{n}\sigma_i)\pi_{n}^{(s,i)}-\frac{\beta_{n-1}\alpha_{n}}{\alpha_{n-1}}\pi_{n-1}^{(s,i)}, \label{eqn:spi}\\
\alpha_{n}^{(i)}&=&\frac{\pi_n^{(s,i)}}{\pi_{n+1}^{(s,i)}}\alpha_{n}, \label{eqn:salpha}\\
\beta_{n-1}^{(i)}&=&\Bigl(\frac{\pi_{n-1}^{(s,i)}}{\pi_{n}^{(s,i)}}\Bigr)^2\beta_{n-1}, \label{eqn:sbeta}\\
x_{n}^{(i)}&=&x_{n-1}^{(i)}+\alpha_{n-1}^{(i)}p_{n-1}^{(i)}, \\
p_{n}^{(i)}&=&r_{n}^{(i)}+\beta_{n-1}^{(i)}p_{n-1}^{(i)},
\label{eqn:endcollinear}
\end{eqnarray}
where $\pi^{(s,i)}_0=\pi^{(s,i)}_{-1}=1$ and the other initial conditions are same as the reference system. Using the information of the reference system, the shifted linear equations $(A+\sigma_i I)x^{(i)}=b$ are readily solved. Note that the computational costs for Eqs.~(\ref{eqn:collinear})-(\ref{eqn:endcollinear}) are negligible because they consist of scalar-scalar and/or scalar-vector products. In addition, $\pi^{(s,i)}_{n-1}$ and $\pi^{(s,i)}_{n}$ as well as the required elements of $x^{(i)}_n$ and $p^{(i)}_n$ for the sampling energy points are stored in memory during the iterations. The iterations are carried out until all norms of the residual vectors satisfy the convergence criterion. 

When the residual vector of the reference system is small, the numerical precision of the residual vectors of the sampling energy points degrades. The seed switching technique replaces the reference system by the energy point having the largest norm of the residual vector among the sampling energy points to circumvent this numerical difficulty. It is obvious from Eqs.~(\ref{eqn:spi})-(\ref{eqn:endcollinear}) that $\pi_n^{(s,i)}$ and $\pi_{n+1}^{(s,i)}$ have to be updated after the reference system changes while $\alpha_{n}^{(i)}$, $\beta_{n}^{(i)}$, $x_n^{(i)}$, and $p_n^{(i)}$ are independent from the choice of the reference system. From the collinear relation of the residual vectors in Eq.~(\ref{eqn:collinear}), $\pi_{n-1}^{(\tilde{s},i)}$ and $\pi_{n}^{(\tilde{s},i)}$ are derived as
\begin{eqnarray}
\pi_{k}^{(\tilde{s},i)}&=&\frac{\pi_{k}^{(s,i)}}{\pi_{k}^{(s,\tilde{s})}}, \:\: (k=n-1 \:\: \mbox{and} \:\: n) 
\label{eqn:seedswitch}
\end{eqnarray}
where $\tilde{s}$ is the index of the new reference system. The extension to the switching to an arbitrary energy point $\sigma$($\notin \{\sigma_1, \sigma_2,\ldots, \sigma_m\}$) is also available \cite{ss}.

\end{document}